\documentclass[conference]{IEEEtran}

\pdfoutput=1 

\IEEEoverridecommandlockouts 

\usepackage[cmex10]{amsmath}
\usepackage{graphicx}
\usepackage{cite}
\usepackage{array}
\usepackage{url}
\usepackage{cases}
\usepackage{amsfonts,amssymb,amsthm}
\usepackage{booktabs}

\graphicspath{{./figures/}}

\DeclareMathAlphabet{\CMmathcal}{OMS}{cmsy}{m}{n}
\renewcommand{\mathcal}[1]{\CMmathcal{#1}}
\newcommand{\figref}[1]{Fig.~\ref{#1}}

\usepackage{cite}
\usepackage{setspace}
\newcommand{\comment}[1]{{{}}}

\newcommand{\eqtop}[1]{\overset{(#1)}{=}}

\newcommand{\vect}[1]{\mathbf{#1}}

\newtheorem{proposition}{Proposition}

\begin{document}

\title{Energy Efficiency and Sum Rate when Massive MIMO meets Device-to-Device Communication}

\author{\IEEEauthorblockN{Serveh~Shalmashi\IEEEauthorrefmark{1}, Emil~Bj\"ornson\IEEEauthorrefmark{2}, Marios~Kountouris\IEEEauthorrefmark{3},  Ki~Won~Sung\IEEEauthorrefmark{1}, and M\'erouane~Debbah\IEEEauthorrefmark{3} } %

\IEEEauthorblockA{\IEEEauthorrefmark{1}Dept.\ of Communication Systems, KTH Royal Institute of Technology, Stockholm, Sweden}
\IEEEauthorblockA{\IEEEauthorrefmark{2}Dept.\ of Electrical Engineering (ISY), Link\"oping University, Link\"oping, Sweden}
\IEEEauthorblockA{\IEEEauthorrefmark{3}Mathematical and Algorithmic Sciences Lab, France Research Center, Huawei Technologies Co. Ltd.} %

Emails: \{serveh,sungkw\}@kth.se, emil.bjornson@liu.se, \{marios.kountouris,merouane.debbah\}@huawei.com
}

\maketitle

\begin{abstract}
This paper considers a scenario of short-range communication, known as device-to-device (D2D) communication, where D2D users reuse the downlink resources of a cellular network to transmit directly to their corresponding receivers. In addition, multiple antennas at the base station (BS) are used in order to simultaneously support multiple cellular users using multiuser or massive MIMO. The network model considers a fixed number of cellular users and that D2D users are distributed according to a homogeneous Poisson point process (PPP). Two metrics are studied, namely, average sum rate (ASR) and energy efficiency (EE). We derive tractable expressions and study the tradeoffs between the ASR and EE as functions of the number of BS antennas and density of D2D users for a given coverage area.
\end{abstract}

\section{Introduction}

\bstctlcite{IEEEexample:BSTcontrol} 

The research on future mobile broadband networks, referred to as the fifth generation (5G), has started in the past few years. In particular, stringent key performance indicators (KPIs) and tight requirements have been introduced in order to handle higher mobile data volumes, reduce latency, increase connectivity and at the same time increase energy efficiency (EE) \cite{Osseiran-2014-COMM, Bjornson-2014-b-SPM}. The current network and infrastructure cannot cope with 5G requirements---fundamental changes are needed to handle future non-homogeneous networks as well as new trends in user behavior such as high quality video streaming and future applications like augmented reality. 5G technology is supposed to evolve existing networks and at the same time integrate new dedicated solutions to meet the KPIs \cite{Bjornson-2014-b-SPM}. The new key concepts for 5G include massive MIMO (multiple-input multiple-output), ultra dense networks (UDN), device-to-device (D2D) communications, and huge number of connected devices, known as machine-type communications (MTC). The potential gains and properties of these different solutions have been studied individually, but the practical gains when they coexist and share network resources are not very clear so far. In this paper, we study the coexistence of two main concepts, namely massive MIMO and D2D communication.

Massive MIMO is a type of multiuser MIMO (MU-MIMO) technology where the base station (BS) uses an array with hundreds of active antennas to serve tens of users on the same time/frequency resources by coherent precoding\cite{Marzetta2010a,Rusek-2013-SPM}. Massive MIMO techniques are particularly known to be very spectral efficient, in the sense of delivering sum rates \cite{Bjornson2016a}. This comes at the price of more hardware, but the solution is still likely to be energy efficient \cite{Ngo2013a,Bjornson-2014-arxiv}.
On the other hand, in a D2D communication, user devices can communicate directly with each other and the user plane data is not sent through the BS \cite{Doppler-2009-COMM}. D2D users either share the resources with cellular networks (overlay approach) or reuse the same spectrum (underlay approach). D2D communication is used for close proximity applications and can bring tremendous gains in terms of offloading core networks and achieving higher data rates with less transmission energy.

We consider two network performance metrics in this work: The average sum rate (ASR) in $\mathrm{bit/s}$ and the EE defined as the number of bits transmitted per Joule of energy consumed by the transmitted signals and the transceiver hardware. It is well known that these metrics depend on the network infrastructure, radio interface, and underlying system assumptions \cite{Tombaz-2011-WCM, Bjornson-2014-arxiv, Auer-2013-WCM}.
The motivation behind our work is to study how the additional degrees of freedom resulting from high number of antennas in the BS can affect the ASR and EE of a multi-tier network where a D2D tier is bypassing the BS. We focus on the downlink since majority of the payload data and network energy consumption are coupled to the downlink \cite{Tombaz-2011-WCM}. We assume that each D2D pair is transmitting simultaneously with the BS in an underlay fashion. In addition, we assume that the communication mode of each user (i.e., D2D or cellular mode) has already been decided by higher layers.

\subsection{Related Work}

The relation between the number of BS antennas, ASR and EE in cellular networks has been studied in \cite{Ngo2013a, Yang-2013-OnlineGreenCom, Bjornson-2014-arxiv, Bjornson-2013-ICT} among others. The tradeoff between ASR and EE was described in \cite{Ngo2013a} for massive MIMO systems with negligible circuit power consumption. This work was continued in \cite{Yang-2013-OnlineGreenCom} where radiated power and circuit power were considered. In \cite{Bjornson-2014-arxiv}, joint downlink and uplink design of a network is studied in order to maximize EE for a given coverage area. The maximal EE was achieved by having a hundred BS antennas and serving tens of users in parallel, which matches well with the massive MIMO concept. Furthermore, the study \cite{Bjornson-2013-ICT} considered a downlink scenario in which a cellular network has been overlaid by small cells. It was shown that by increasing the number of BS antennas, the array gain allows for decreasing the radiated signal energy while maintaining the same ASR. However, the energy consumed by circuits increases. Maximizing the EE is thus a complicated problem where several counteracting factors need to be balanced. This stands in contrast to maximization of the ASR, which is rather straightforward since the sum capacity is the fundamental upper bound.

There are only a few works in the D2D communication literature where the base stations have multiple antennas \cite{Min-2011-TWC-b, Yu-2012-GLOBECOM, Fodor-2011-GLOBECOM,Shalmashi-2014a-WCNC, Xingqin-2014-arxiv}.
In \cite{Min-2011-TWC-b}, uplink MU-MIMO with one D2D pair was considered. Cellular user equipments (CUEs) were scheduled if they are not in the interference-limited zone of the D2D user. The study \cite{Yu-2012-GLOBECOM} compared different multi-antenna transmission schemes. In \cite{Fodor-2011-GLOBECOM}, two power control schemes were derived in a MIMO network. Two works that are more related to our work are \cite{Shalmashi-2014a-WCNC} and \cite{Xingqin-2014-arxiv}. The former investigates the mode selection problem in the uplink of a network with potentially many antennas at the BS. The impact of the number of antennas on the quality-of-service and transmit power was studied while users need to decide their mode of operation (i.e., D2D or cellular). The latter study, \cite{Xingqin-2014-arxiv}, only employs extra antennas in the network to protect the CUEs in the uplink.

The ASR in D2D communications is mostly studied in the context of interference and radio resource management \cite{Shalmashi-2013-PIMRC, Zulhasnine-2010-WiMob}. There are a few works that consider EE in D2D communications, but only for single antenna BSs, e.g.,  \cite{Yaacoub-2012-GCW, Mumtaz-2014-ICC}, and \cite{Wang-2013-ICC}, where the first one proposed a coalition formation method, the second one  designed a resource allocation scheme, and the third one aimed at optimizing the battery life of user devices.

The degrees of freedom offered by having multiple antennas at BSs are very useful in the design of future mobile networks, because the spatial precoding enables dense multiplexing of users with little inter-user interference. In particular, the performance for cell edge users, which have almost equal SNR to several BSs, can be greatly improved since only the useful signal are amplified by  precoding \cite{Baldemair-2013-VTM,Bjornson2013d,Gesbert-2007-SPM}. In order to model random user positions, we use mathematical tools from stochastic geometry \cite{Haenggi-Stochastic}. We assume that there is a fixed number of randomly distributed  CUEs in the network, while the locations of D2D transmitters are randomly distributed according to a homogeneous Poisson point process (PPP). We characterize the relation between the ASR and EE metrics in terms of number of antennas and D2D user density with fixed number of CUEs for a given coverage area.

\section{System Model}
\label{sec:Sys_Mod}

\begin{figure}[t]
\centering
\includegraphics[width=0.85\columnwidth]{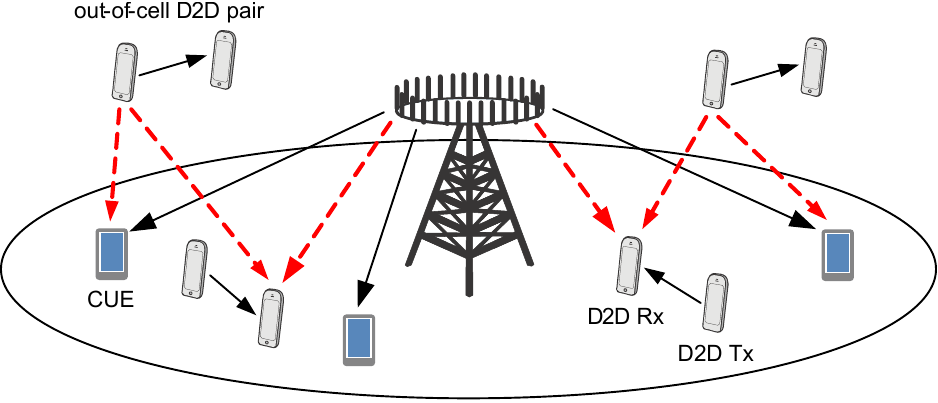}
\vspace*{-3mm}
\caption{System model where a multi-antenna BS communicates in the downlink with multiple CUEs, while multiple user pairs communicate in D2D mode. The CUEs are distributed uniformly in the coverage area and the D2D users are distributed according to a PPP. The D2D users that are outside the coverage area are only considered as interferers.}
\label{figure:sysMod_massiveMIMO_D2D}
\vspace*{-6mm}
\end{figure}

We consider a single-cell scenario where the BS is located in the center of the cell and its coverage area is a disc of radius $R$. The BS serves $U_c$ single-antenna CUEs which are uniformly distributed in the coverage area. These are simultaneously served in the downlink using an array of $T_c$ antennas located at the BS. It is assumed that $1 \leq U_c \leq T_c$ so that the beamforming can control interference \cite{Bjornson-2014-a-SPM}.

In addition to the CUEs, there are other single-antenna users that bypass the BS and communicate pairwise with each other using a D2D communication mode. The locations of the D2D transmitters (D2D Tx) are modeled by a homogeneous Poisson point process (PPP) $\Phi$ with density $\lambda_d$ in $\mathbb{R}^2$, i.e., the average number of D2D Tx per unit area is $\lambda_d$. The D2D receiver (D2D Rx) is randomly generated in an isotropic direction with a fixed distance away from its corresponding D2D Tx similar to the model considered in \cite{Lee-2014-JSAC}. The system model is shown in Fig.~\ref{figure:sysMod_massiveMIMO_D2D}. Let $R_{k,j}$ denote the distance between the $j$-th D2D~Tx to the $k$-th D2D~Rx. %
The performance analysis for D2D users is carried out for a typical D2D user, which is denoted by the index  $0$. The typical D2D user is an arbitrarily D2D user located in the cell and its corresponding receiver is positioned in the origin. The results for a typical user show the statistical average performance of the network \cite{Haenggi-Stochastic}. Therefore, for any performance metric derivation, the D2D users inside the cell are considered and the ones outside the cell are only taken into account as sources of interference. Note that we neglect potential interference from other BSs and leave the multicell case for future work. We assume equal power allocation for both CUEs and D2D users. Let $P_c$ denote the total transmit power of the BS, then the transmit power per CUE is $\frac{P_c}{U_c}$. The transmit power of the D2D Tx is denoted by $P_d$.

Let $\vect{h}_j \in \mathbb{C}^{T_c \times 1}$ be the normalized channel response between the BS and the $j$-th CUE, for $j \in \{0, \dots, U_c-1 \}$. These channels are modeled as Rayleigh fading such that $\vect{h}_j \sim\mathcal{CN}(\mathbf{0},\mathbf{I})$, where $\mathcal{CN}(\cdot,\cdot)$ denotes a circularly symmetric complex Gaussian distribution. Linear downlink precoding is considered at the BS, based on the zero-forcing (ZF) scheme that mitigates interference between the CUEs \cite{Bjornson-2014-a-SPM}. The precoding matrix is denoted by $\mathbf{V}=[\vect{v}_0, \dots, \vect{v}_{Uc-1}]\in \mathbb{C}^{T_c \times U_c}$ in which each column $\vect{v}_j$ is a normalized precoding vector that is assigned to the CUE $j$ in the downlink. Let $\vect{f}_{0,\textrm{BS}} \in \mathbb{C}^{T_c \times 1}$ be the channel response from the BS to D2D Rx and let it be Rayleigh fading as  $\vect{f}_{0,\textrm{BS}} \sim \mathcal{CN}(\mathbf{0},\mathbf{I})$.
Moreover, let $r_j \in \mathbb{C}$ and $\vect{s} \in \mathbb{C}^{U_c \times 1}$ denote the transmitted data signals intended for a D2D Rx and the CUEs, respectively. Since each user requests a different data, the transmitted signals can be modeled as zero-mean and uncorrelated with $\mathbb{E}\big[|r_j|^2\big]=P_d$ and $\mathbb{E}\big[||\vect{s}||^2\big]=P_c$.
The fading channel response between the $j$-th D2D~Tx and the $k$-th D2D~Rx is denoted by $g_{k,j} \in \mathbb{C}$ where $g_{k,j} \sim\mathcal{CN}(0,1)$. Moreover, $R_{0,\textrm{BS}}$ denotes the random distance between the typical D2D Rx and the BS. The pathloss is modeled as $A_{0,i}d^{-\alpha_i}$ with $i \in \{c,d\}$, where index $c$ indicates the pathloss between a user and the BS and index $d$ corresponds to the pathloss between any two users. $A_{0,i}$ and $\alpha_i$  are pathloss coefficient and exponent, respectively. Perfect instantaneous channel state information (CSI) is assumed in this work for analytic tractability, but imperfect CSI is a relevant extension. %
The received signal at the typical D2D Rx is
\begin{align}
y_{d,0} &= \sqrt{A_{0,d}} R_{0,0}^{-\alpha_d/2} g_{0,0} r_0 +\underbrace{ \sqrt{A_{0,c}} R_{0,\textrm{BS}}^{-\alpha_c/2} \vect{f}_{0,\textrm{BS}}^H \mathbf{V} \vect{s} }_{\textrm{Interference from the BS}} \nonumber\\
&\quad + \underbrace{\sqrt{A_{0,d}} \sum_{j\neq 0} R_{0,j}^{-\alpha_d/2} g_{0,j} r_j}_{\textrm{Interference from other D2D users}} + \eta_{d,0},
\end{align}
where $\eta_{d,0}$ is zero-mean additive white Gaussian noise with  power  $N_0 = \tilde{N_0}B_w$, $\tilde{N_0}$ is the power spectral density of the white Gaussian noise, and $B_w$ is the channel bandwidth. %
Therefore, the signal-to-interference-plus-noise ratio (SINR) at the typical D2D Rx is
\begin{align}\label{eq: SINR_d2d}
\textrm{SINR}_{d,0} = \frac{P_d  R_{0,0}^{-\alpha_d}|g_{0,0}|^2}{I_{\textrm{BS},0} + I_{d,0} +  \frac{N_0}{A_{0,d}}},
\end{align}
in which both numerator and denominator have been normalized by $A_{0,d}$, and
\begin{align}
I_{\textrm{BS},0} &=\frac{ A_{0,c}}{A_{0,d}}\frac{P_c}{U_c}R_{0,\textrm{BS}}^{-\alpha_c}||\vect{f}_{0,\textrm{BS}}^H \mathbf{V}||^2, \label{eq: I_BS0} \\
I_{d,0} &= \sum_{j\neq 0} P_d R_{0,j}^{-\alpha_d} |g_{0,j}|^2, \label{eq: I_d0}
\end{align}
where $I_{\textrm{BS},0}$ is the received interference power from the BS and $I_{d,0}$ is the received interference power from other D2D users that transmit simultaneously.

Let $D_{0,k}$ and $e_{0,k}\in \mathbb{C}$ with $e_{0,k}\sim\mathcal{CN}(0,1)$  be the distance and fading channel response between a typical CUE and the $k$-th D2D Tx, respectively, and let $D_{0,\textrm{BS}}$ denote the distance between a typical CUE and the BS. Then, the received signal at the typical CUE is
\begin{align}
y_{c,0} &= \sqrt{A_{0,c}} D_{0,\textrm{BS}}^{-\alpha_c/2} \vect{h}_{0}^H \mathbf{V} \vect{s}  \notag \\
&\qquad + \underbrace{\sqrt{A_{0,d}} \sum_{j} D_{0,j}^{-\alpha_d/2} e_{0,j} r_j}_{\textrm{Interference from all D2D users}} + \eta_{c,0},
\end{align}
where $\eta_{c,0}$  is zero-mean additive white Gaussian noise with power $N_0$. Using the notation $\zeta = A_{0,c}\frac{P_c}{U_c}$, the corresponding SINR for the typical CUE is
\begin{align}
\textrm{SINR}_{c,0} = \frac{|\vect{h}_{0}^H \vect{v}_{0}|^2}{\frac{A_{0,d}}{\zeta} D_{0,\textrm{BS}}^{\alpha_c}(I_{d,c} + \frac{N_0}{A_{0,d}})}, \label{eq: SINR_CUE}
\end{align}
where
\begin{equation}\label{eq: inf_dc}
I_{d,c} = \sum_{j} P_d D_{0,j}^{-\alpha_d} |e_{0,j}|^2
\end{equation}
is the received interference power from all D2D users.

\section{Performance Analysis}

\subsection{Metrics}
\label{sec:metrics}

In this paper, two main performance metrics for the network are considered:  the average sum rate (ASR) and energy efficiency (EE). %
The ASR is obtained from total rates of both D2D and cellular users as
\begin{equation}
\text{ASR}= U_c \bar{R}_c +  \pi R^2 \lambda_d \bar{R}_d,
\label{eq:ASR}
\end{equation}
where $\pi R^2 \lambda_d$ is the average number of D2D users in the cell and $\bar{R}_t$ with $t\in\{c,d\}$ denotes the average rates of the CUEs and D2D users, respectively. To compute the ASR, we use the following lower bound on the average rate for both cellular and D2D users \cite[Corollary~2]{Ye-2014-ICC}
\begin{equation}\label{eq:Avg_Rate}
\bar{R}_t = \underset {\beta_t \geq 0}{\mathrm{sup}}~B_w \log_{2}\left(1 + \beta_t \right) \mathrm{Pr}\big\{\mathrm{SINR}_{t} \geq \beta_t \big\},
\end{equation}
which corresponds to the successful transmission rate, and
\begin{equation}\label{eq:PsuccDef}
\mathrm{P_{succ}} = \mathrm{Pr}\big\{\mathrm{SINR}_{t} \geq \beta_t \big\}
\end{equation}
is the probability of coverage when the received SINR is higher than a specified threshold $\beta_t$ needed for successful reception. Note that $\mathrm{SINR}_{t}$ contains random fading and random user locations. Finding the supremum guarantees the best rate for the D2D user or CUE.  If we know the probability of coverage, the expression in \eqref{eq:Avg_Rate} can easily be computed by using line search for each user type independently.

The EE is defined as the ratio between the ASR and the total consumed power, i.e.,
\begin{equation}
\text{EE} = \frac{\text{ASR}}{\text{Total power}}. \label{eq:EE}
\end{equation}
For the total power consumption, we consider a detailed  model described in \cite{Bjornson-2014-arxiv} as
\begin{align}
\text{Total power}&=\frac{1}{\eta} \big(P_c +\lambda_d \pi R^2 P_d\big) + C_0  \notag\\
&\;\quad + T_c C_1 + \big(U_c + 2 \lambda_d \pi R^2\big)C_2,
\end{align}
where $P_c +\lambda_d \pi R^2 P_d$ is the total transmission power, $\eta$ is the amplifier efficiency ($0 < \eta \leq 1$), $C_0$ is the load independent power consumption at the BS, $C_1$ is the power consumption per BS antennas, $C_2$ is the power consumption per handset, and $U_c + 2 \lambda_d \pi R^2$ is the number of active users.

\subsection{Probability of Coverage}
In order to calculate the ASR and EE, we derive the probability of coverage for both cellular and D2D users. This is one of the main contributions of the paper. Due to the page limit, we only provide outlines of the proofs.
\begin{proposition} \label{proposition:P_succ_d2d}
The probability of coverage for a typical D2D user is given by
\begin{align}
\mathrm{P^{{d2d}}_{succ}} &=    \frac{ \kappa(x)^{2/\alpha_c}}{R^2}\Bigg[ y^{U_c + \frac{2}{\alpha_c}-1} (1-y)^{- \frac{2}{\alpha_c}}  \notag\\
&\;\quad  - \big(U_c + \frac{2}{\alpha_c}-1\big) \mathcal{B}\big( y; U_c + \frac{2}{\alpha_c}-1, 1-\frac{2}{\alpha_c}\big) \Bigg] \notag \\
&\;\quad \cdot \exp\left(- \frac{\pi \lambda_d R_{0,0}^2}{\mathrm{sinc}(\frac{2}{\alpha_d})} \beta_d^{\frac{2}{\alpha_d}}\right) \exp\left(- \beta_d x\frac{N_0}{A_{0,d}}\right),
\label{eq:P_succ_d2d}
\end{align}
where $x \triangleq \frac{R_{0,0}^{\alpha_d}}{P_d}$, $\kappa(x) \triangleq  \beta_d x \frac{ A_{0,c}}{ A_{0,d} }\frac{P_c}{U_c}$, $y \triangleq \frac{1}{\kappa (x) R^{-\alpha_c} + 1}$, $\mathrm{sinc}(z) = \frac{\sin(\pi z)}{\pi z}$, and $\mathcal{B}(y;a,b)$ is the incomplete Beta function.
\end{proposition}
\begin{IEEEproof}
This result follows by substituting the definition of $\textrm{SINR}_{d,0}$ from \eqref{eq: SINR_d2d} into \eqref{eq:PsuccDef} %
where we obtain
\begin{align}
\mathrm{P^{\mathrm{d2d}}_{succ}} &= \mathrm{Pr} \Big\{ |g_{0,0}|^2 \geq  \frac{R_{0,0}^{\alpha_d}}{P_d}  \beta_d \big(I_{\mathrm{BS},0} + I_{d,0} +  \frac{N_0}{A_{0,d}}\big) \Big\} \nonumber \\
&\eqtop{a} \mathbb{E}_{I_{\mathrm{BS},0}, I_{d,0}} \Big[e^{- \beta_d x (I_{\mathrm{BS},0} + I_{d,0} +  \frac{N_0}{A_{0,d}})}\Big]\nonumber \\
&\eqtop{b}     \mathcal{L}_{I_{\mathrm{BS},0}}(\beta_d x)\mathcal{L}_{I_{d,0}}(\beta_d x) e^{- \beta_d x\frac{N_0}{A_{0,d}}}, \label{eq:Pd2dProof}
\end{align}
where $\mathcal{L}$ denotes the Laplace transform %
defined as $\mathcal{L}_{z}(s) = \mathbb{E}_{z}\big[e^{- sz}\big]$. Step $(a)$ comes from the fact that $|g_{0,0}|^2 \sim \exp(1)$ and $(b)$ follows since the interference terms are independent and the noise term is not dependent on the interference terms.  The first Laplace transform in \eqref{eq:Pd2dProof} is with respect to $I_{\mathrm{BS},0}$ in \eqref{eq: I_BS0} which is a function of two random variables, i.e.,  $\|\vect{f}_{0,\textrm{BS}}^H \mathbf{V}\|^2$ where the norm is well-approximated by a Chi-squared distribution  as $2\|\vect{f}_{0,\textrm{BS}}^H \mathbf{V}\|^2 \sim \chi^2_{2U_c}$ \cite{Dhillon-2013-TWC}, %
and $R_{0,\textrm{BS}}$ which is uniformly distributed over the coverage area. Calculating this Laplace transform results in the non-exponential term of \eqref{eq:P_succ_d2d}. The second Laplace transform in \eqref{eq:Pd2dProof} is with respect to $I_{d,0}$ in \eqref{eq: I_d0} and is calculated based on the probability generating functional (PGFL) \cite{Haenggi-Stochastic} which results in the first exponential term in \eqref{eq:P_succ_d2d}.
\end{IEEEproof}

\begin{proposition} \label{proposition:P_succ_cue}
The probability of coverage for a typical cellular user is given by
\begin{align}
\mathrm{P^c_{succ}} &= \mathbb{E}_{D_{0,\textrm{BS}}}\Bigg[e^{-s(\frac{N_0}{A_{0,d}})} \sum_{k=0}^{T_c - U_c} \frac{s^k}{k!} \sum_{i=0}^{k} \binom{k}{i} \Big(\frac{N_0}{A_{0,d}}\Big)^{k-i} \notag\\
&\;\;\qquad \cdot (-1)^i\frac{\mathrm{d}^i}{{\mathrm{d}s}^i} \exp\left(-  \frac{\pi \lambda_d}{\mathrm{sinc}(\frac{2}{\alpha_d})}(s P_d)^{\frac{2}{\alpha_d}}\right)
\Bigg], \label{eq:P_succ_cue}
\end{align}
where $s\triangleq \frac{\beta_c}{\zeta} A_{0,d} D_{0,\textrm{BS}}^{\alpha_c}$ and $\frac{\mathrm{d}^i}{{\mathrm{d}s}^i}$ stands for the $i^\textrm{th}$ derivative.
\end{proposition}
\begin{IEEEproof}
Substituting $\textrm{SINR}_{c,0}$ into \eqref{eq:PsuccDef}, we get
\begin{align}
&\mathrm{P}^{\textrm{c}}_{\textrm{succ}} = \mathrm{Pr}\left\{ |\vect{h}_{0}^H \vect{v}_{0}|^2 \geq \frac{\beta_c}{\zeta}A_{0,d} D_{0,\textrm{BS}}^{\alpha_c}\Big(I_{d,c} + \frac{N_0}{A_{0,d}}\Big) \right\} \nonumber\\
&\eqtop{a} \mathbb{E}_{D_{0,\textrm{BS}}, I_{d,c}}\Bigg[e^{-s(I_{d,c}+\frac{N_0}{A_{0,d}})} \sum_{k=0}^{T_c - U_c} \frac{s^k}{k!} \Big(I_{d,c}+\frac{N_0}{A_{0,d}}\Big)^{k}  \Bigg] \nonumber\\
\end{align}
where $(a)$ follows from the CCDF of $|\vect{h}_{0}^H \vect{v}_{0}|^2$ with $2|\vect{h}_{0}^H \vect{v}_{0}|^2 \sim \chi^2_{2(T_c-U_c+1)}$ given $D_{0,\textrm{BS}}$ and $I_{d,c}$. Taking the expectation with regard to the interference $I_{d,c}$ gives \eqref{eq:P_succ_cue}.
\end{IEEEproof}
The expectation in  \eqref{eq:P_succ_cue} with respect to $D_{0,\textrm{BS}}$  can be computed numerically. The analytical results of Proposition~\ref{proposition:P_succ_d2d} and Proposition~\ref{proposition:P_succ_cue} have been verified by Monte-Carlo simulations. A main benefit of the analytic expressions is that they can be computed efficiently, which basically is a prerequisite for the multi-variable system analysis carried out in Section~\ref{sec:results}. Using the results from Proposition~\ref{proposition:P_succ_d2d} and Proposition~\ref{proposition:P_succ_cue}, we proceed to evaluate the network performance in terms of the ASR and EE from \eqref{eq:ASR} and \eqref{eq:EE}, respectively.

\section{Numerical Results}
\label{sec:results}
\begin{table}[t]
\scriptsize
\centering
\caption{System and simulation parameters.}
\label{table:Sim_param}
\vspace{-2mm}
\begin{tabular}{@{} l c c  @{}} \toprule
  Description & Parameter & Value\\ \midrule
  D2D TX power & $P_d$ & $13$ dBm\\
  BS TX power &  $P_c$ & $41$ dBm\\
  Number of CUE &  $U_c$ & $4$\\
  Cell radius & $R$ & $500$ m\\
  Bandwidth & $B_w$  & $20$ MHz\\
  Thermal noise power & $N_0$ & $-131$ dBm\\
  Noise figure in  UE & $F$ & $5$ dB\\
  Carrier frequency & $f_c$  & $2$ GHz\\
  D2D pair distance & $d_{\text{d2d}}$ & $50$ m\\
  Pathloss exponent betw.\ devices & $\alpha_d$  & $3$\\
  Pathloss exponent betw.\ BS--device & $\alpha_c$  & $3.67$\\
  Pathloss coefficient betw.\ devices & $A_d$  & $38.84$ dB\\
  Pathloss coefficient betw.\ BS--device & $A_c$  & $30.55$ dB\\
  Amplifier efficiency & $\eta$  & $0.3$\\
  Load-independent power in BS & $C_0$  & $5$ W\\
  Power per BS antenna & $C_1$  & $0.5$ W\\
  Power per UE handset & $C_2$  & $0.1$ W\\\bottomrule
\end{tabular}
\vspace{-6mm}
\end{table}

In this section, we assess the performance of the setup in Fig.~\ref{figure:sysMod_massiveMIMO_D2D} in terms of ASR and EE using numerical evaluations. Both EE and ASR are functions of three key parameters, namely, the number of BS antennas $T_c$, the density of D2D users $\lambda_d$, and the number of cellular users $U_c$. %
Due to space limitations, we cannot show the individual effect of all parameters. Therefore, we fix the number of CUEs to $U_c = 4$ and study the tradeoffs among other parameters. %
The system and simulation parameters are given in Table~\ref{table:Sim_param}. %

\begin{figure}[t]
\centering
\includegraphics[width=0.9\columnwidth]{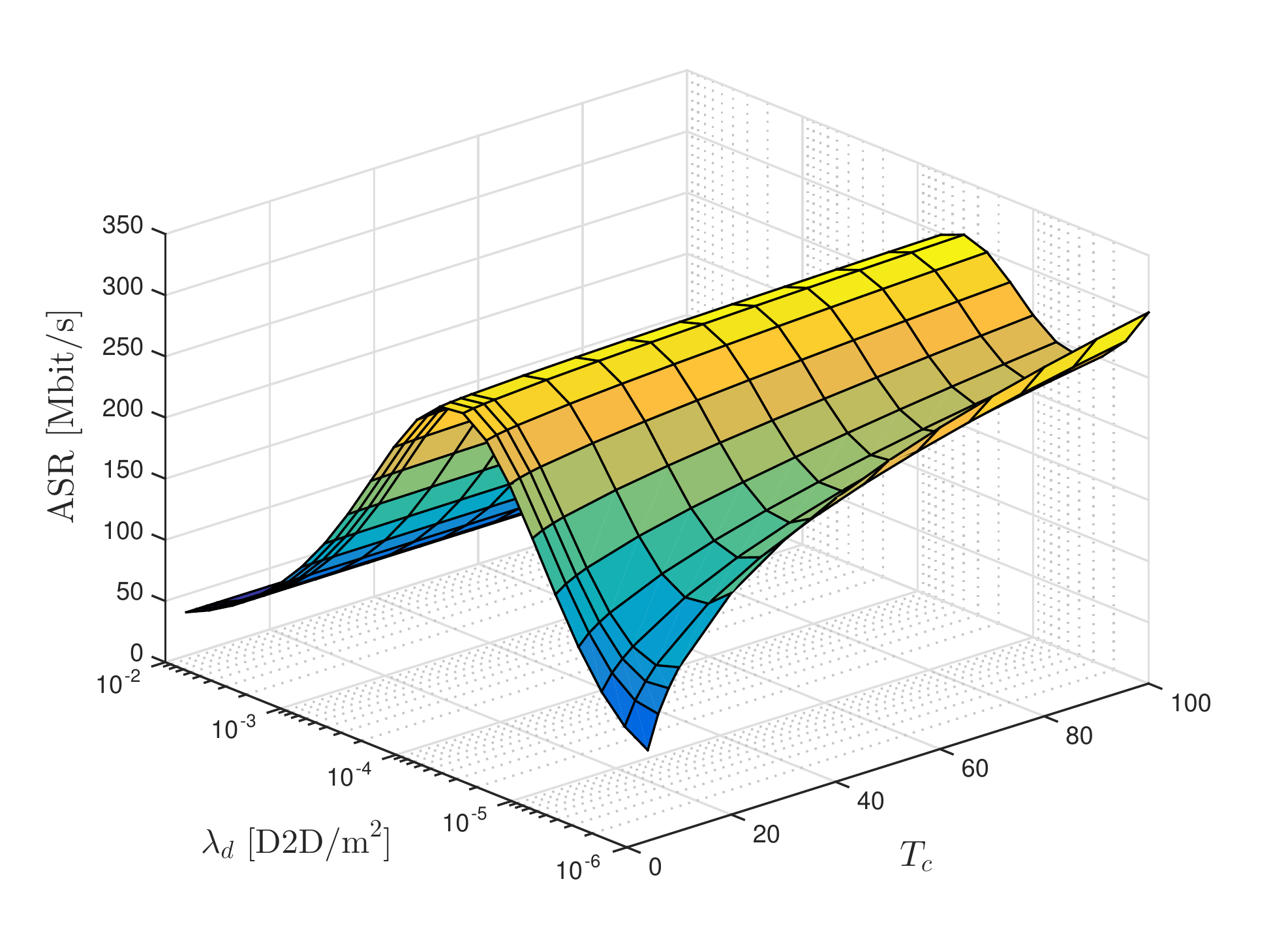}
\vspace*{-3mm}
\caption{ASR $\mathrm{[Mbit/s]}$ as a function of D2D user density $\lambda_d$ and BS antennas $T_c$.}
\label{fig:ASR_ld_tc_uc4}
\vspace*{-4.5mm}
\end{figure}

\begin{figure}[t]
\centering
\includegraphics[width=0.9\columnwidth]{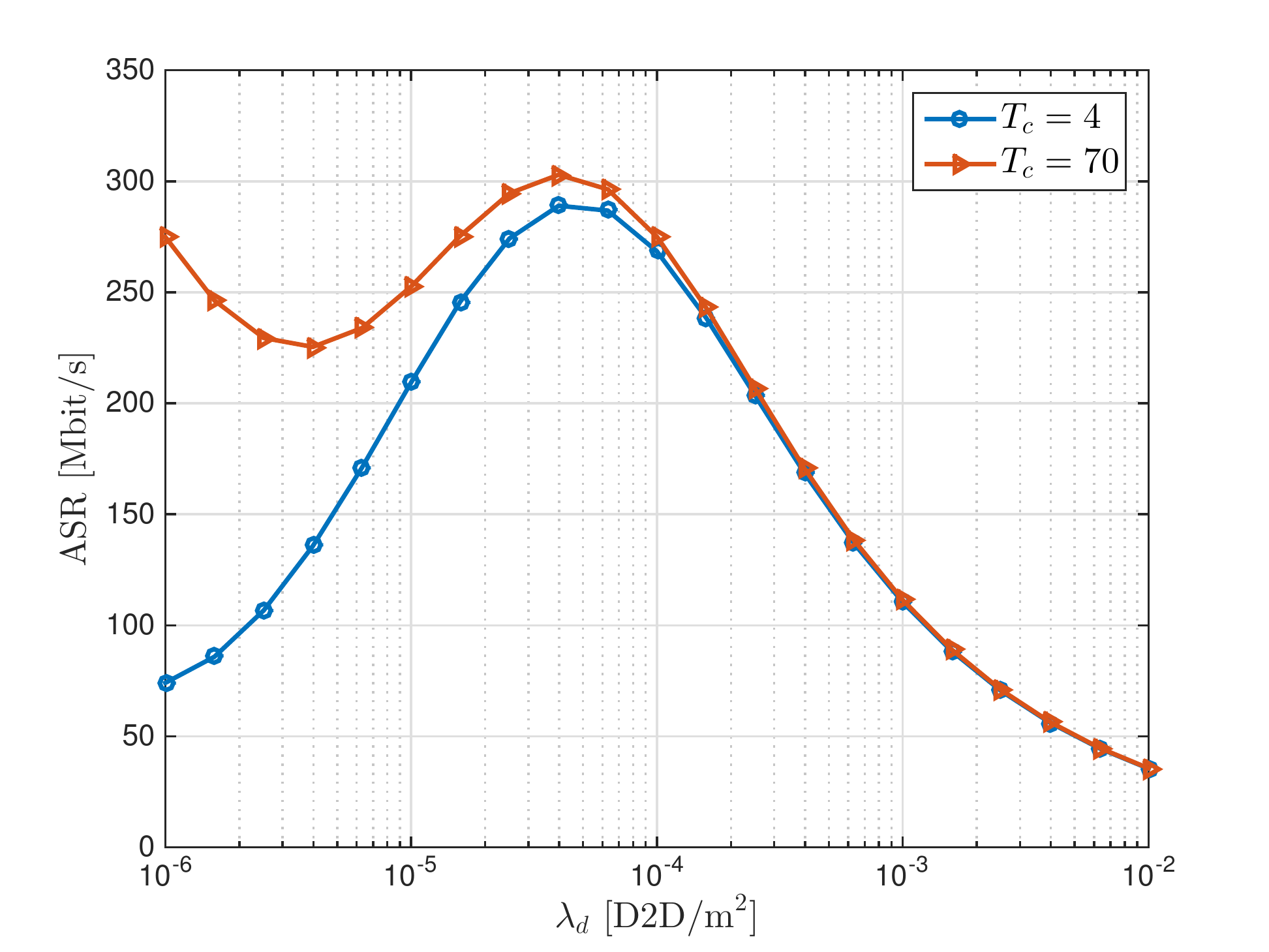}
\vspace*{-3mm}
\caption{ASR $\mathrm{[Mbit/s]}$ as a function of D2D user density $\lambda_d$ for $T_c=\{4,70\}$.}
\label{fig:ASR_ld_tc4_tc70}
\vspace*{-4.5mm}
\end{figure}

In \figref{fig:ASR_ld_tc_uc4}, the behavior of ASR is shown with respect to different number of antennas $T_c$ and the density of D2D users $\lambda_d$. It is observed that increasing the number of antennas always seems to increase the ASR. In contrast, for the ASR, there is an optimal value of the D2D density $\lambda_d$, which results in the maximum ASR for each number of antennas. However, there is a difference in the shape of the ASR in lower $T_c$ values and higher $T_c$ values. In order to clarify this effect, we plot the ASR versus $\lambda_d$ in a 2-D plot with $T_c = \{4,70\}$ in \figref{fig:ASR_ld_tc4_tc70}.

\begin{figure}[t]
\centering
\includegraphics[width=0.9\columnwidth]{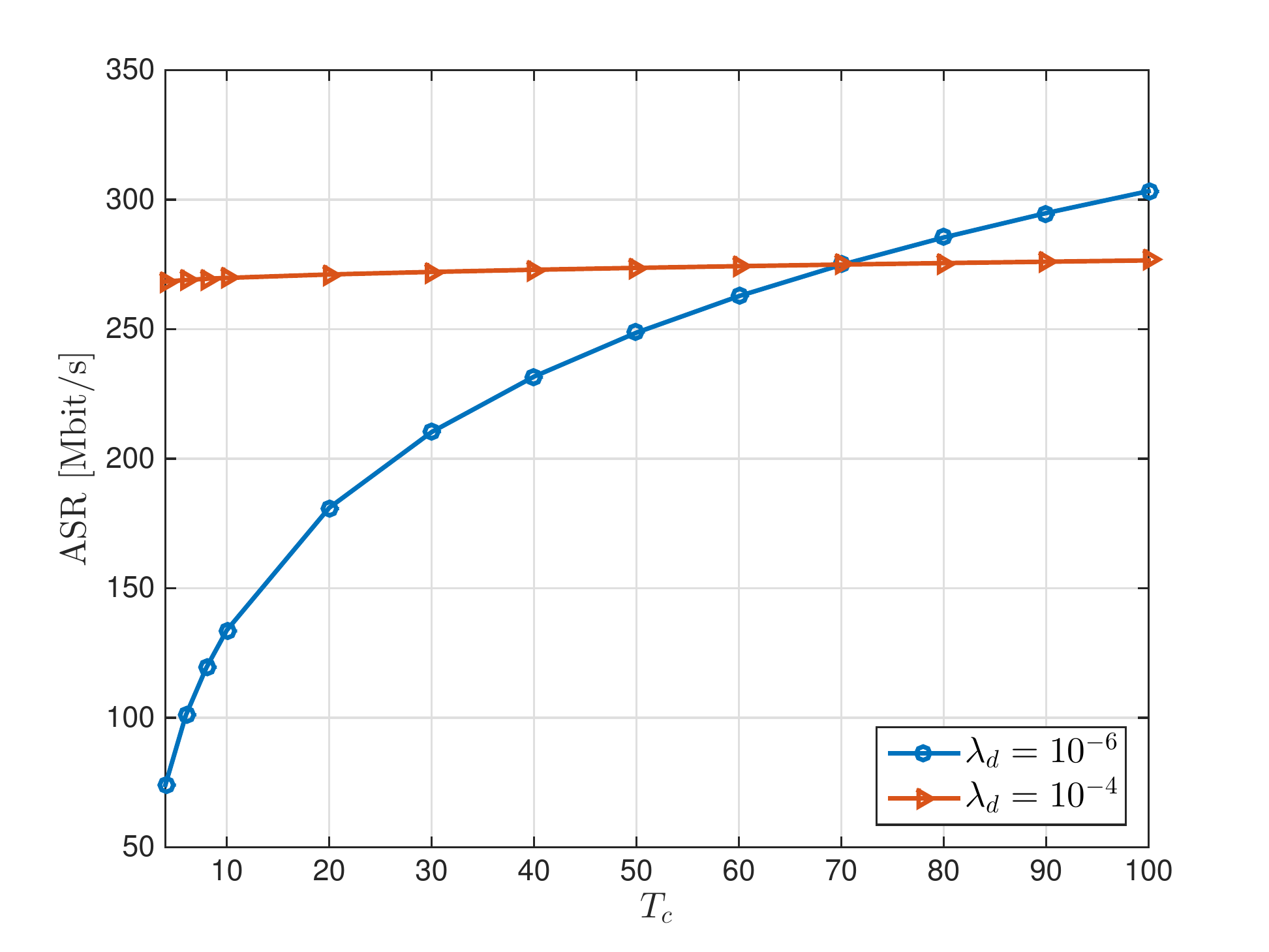}
\vspace*{-3mm}
\caption{ASR $\mathrm{[Mbit/s]}$ as a function of BS antennas $T_c \in \{4,\dots,100\}$ for $\lambda_d=\{10^{-6},10^{-4}\}$.}
\label{fig:ASR_ld-4-6_tc}
\vspace*{-4.5mm}
\end{figure}

For $T_c = 4$, the rate contributed from CUEs to the sum rate is low and adding D2D users to the network (i.e. increasing $\lambda_d$), even though they create interference, leads to increasing ASR until a certain density. This increase in the ASR continues until a point where the cross-tier interference between D2D users limits per link data rate and results in decreasing ASR. By increasing the number of antennas to $T_c = 70$, the rates of the cellular users becomes higher. However, by introducing small number of D2D users, the effect of the initial interference from D2D users becomes visible, i.e. the decrease in the CUEs' rates is not compensated in the ASR by the contribution of the D2D users' rate. However, if we keep increasing the number of D2D users, even though the rate per link decreases for both CUE and D2D users, the resulting aggregate D2D rate becomes higher and the ASR starts to increase, which is the reason behind the local minima. The second turning point follows from the same reasoning as with $T_c = 4$, i.e. in higher D2D densities, the interference from D2D users are the limiting factor for the ASR. This effect can also be observed in \figref{fig:ASR_ld-4-6_tc} where the ASR performance is depicted versus different number of antennas for two D2D densities. At lower densities, increasing the number of antennas is beneficial in terms of ASR, however, in the interference-limited regime (higher $\lambda_d$), increasing the number of antennas does not impact the total network performance.

\begin{figure}[t]
\centering
\includegraphics[width=0.9\columnwidth]{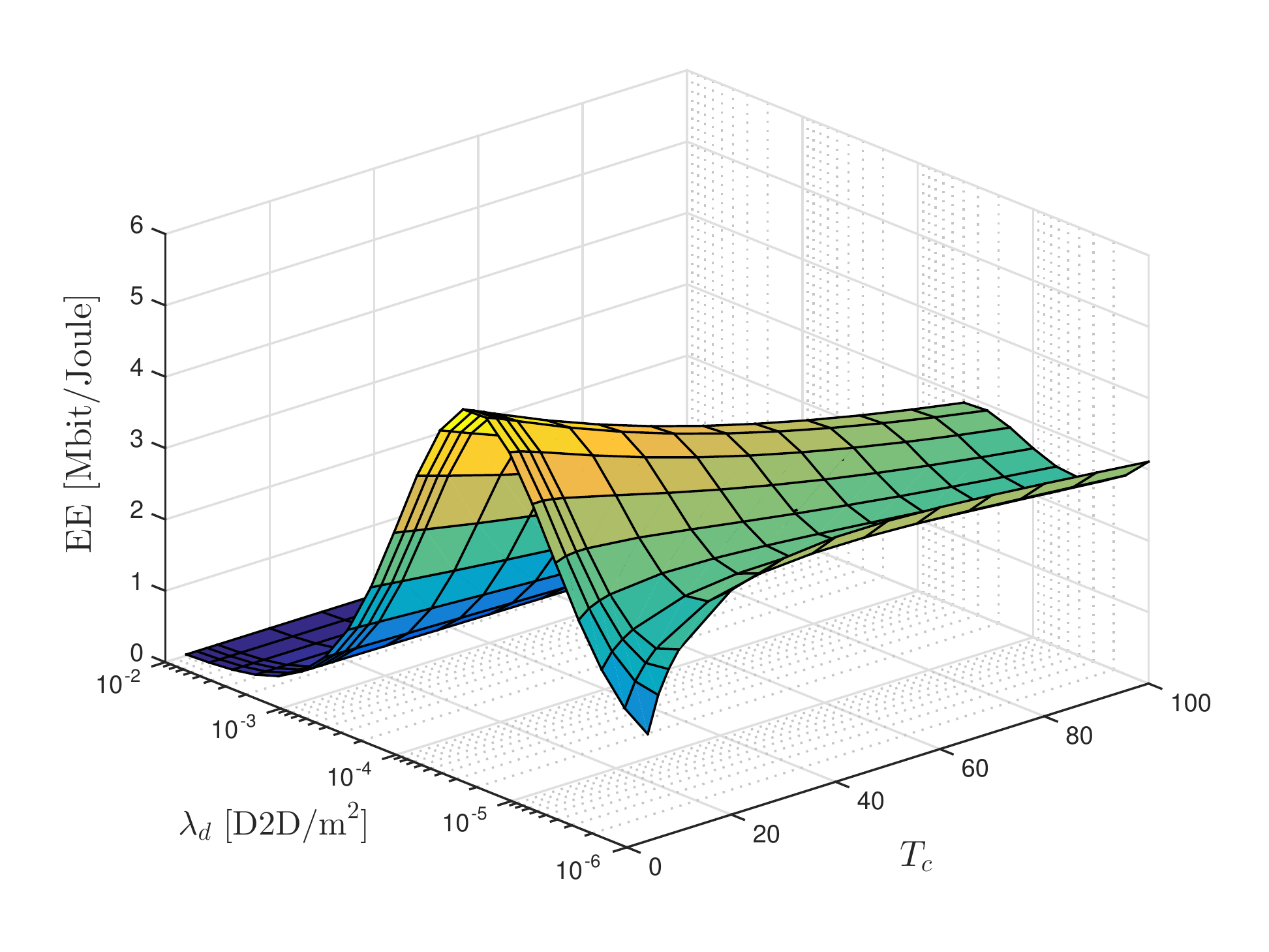}
\vspace*{-3mm}
\caption{EE $\mathrm{[Mbit/Joule]}$ as a function of D2D user density $\lambda_d$ and BS antennas $T_c$.}
\label{fig:EE_ld_tc_uc4}
\vspace*{-4.5mm}
\end{figure}

\begin{figure}[t]
\centering
\includegraphics[width=0.9\columnwidth]{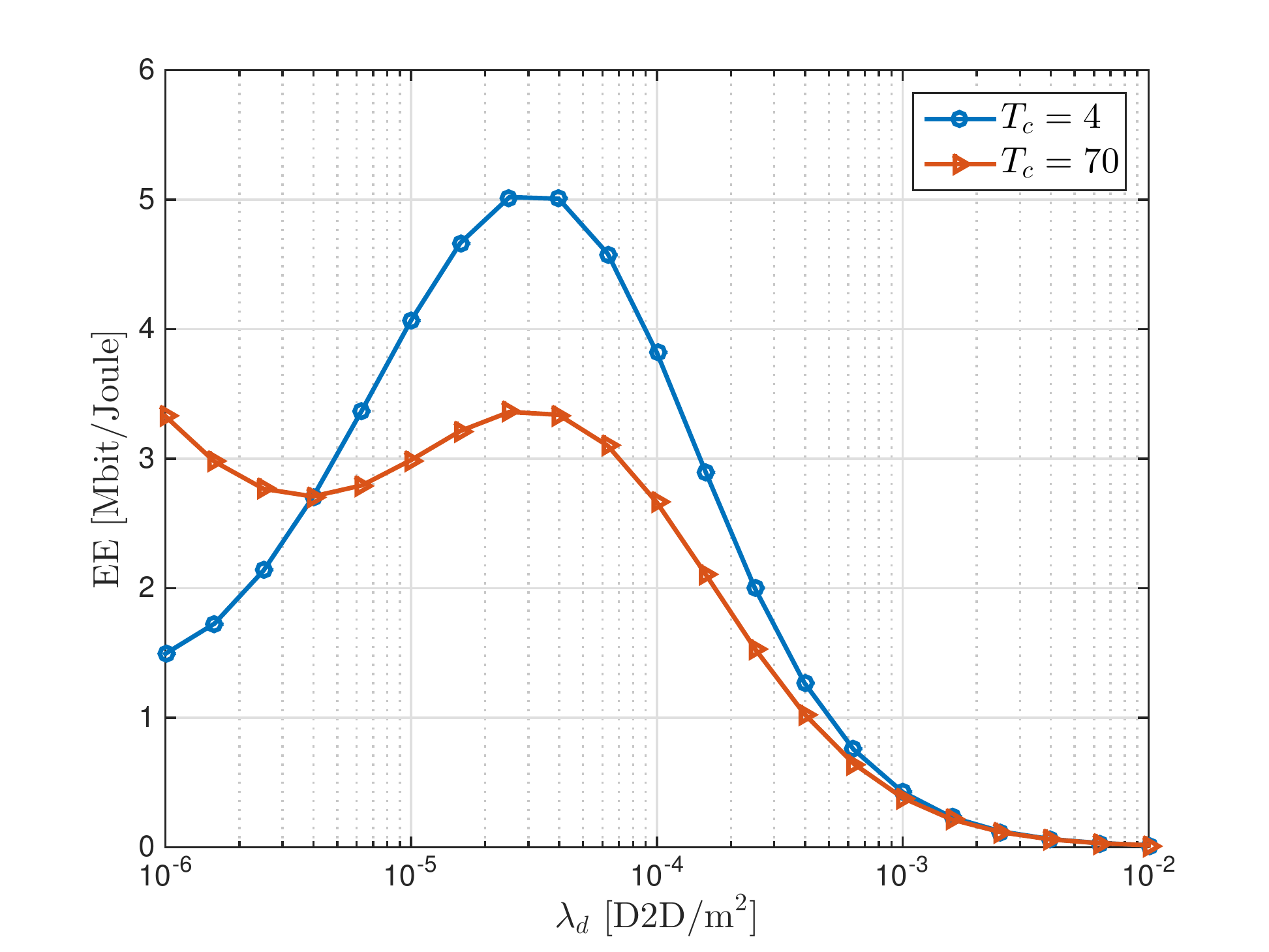}
\vspace*{-3mm}
\caption{EE $\mathrm{[Mbit/Joule]}$ as a function of D2D user density $\lambda_d$ for $T_c=\{4,70\}$.}
\label{fig:EE_ld_tc4_tc70}
\vspace*{-5mm}
\end{figure}

\begin{figure}[t]
\centering
\includegraphics[width=0.9\columnwidth]{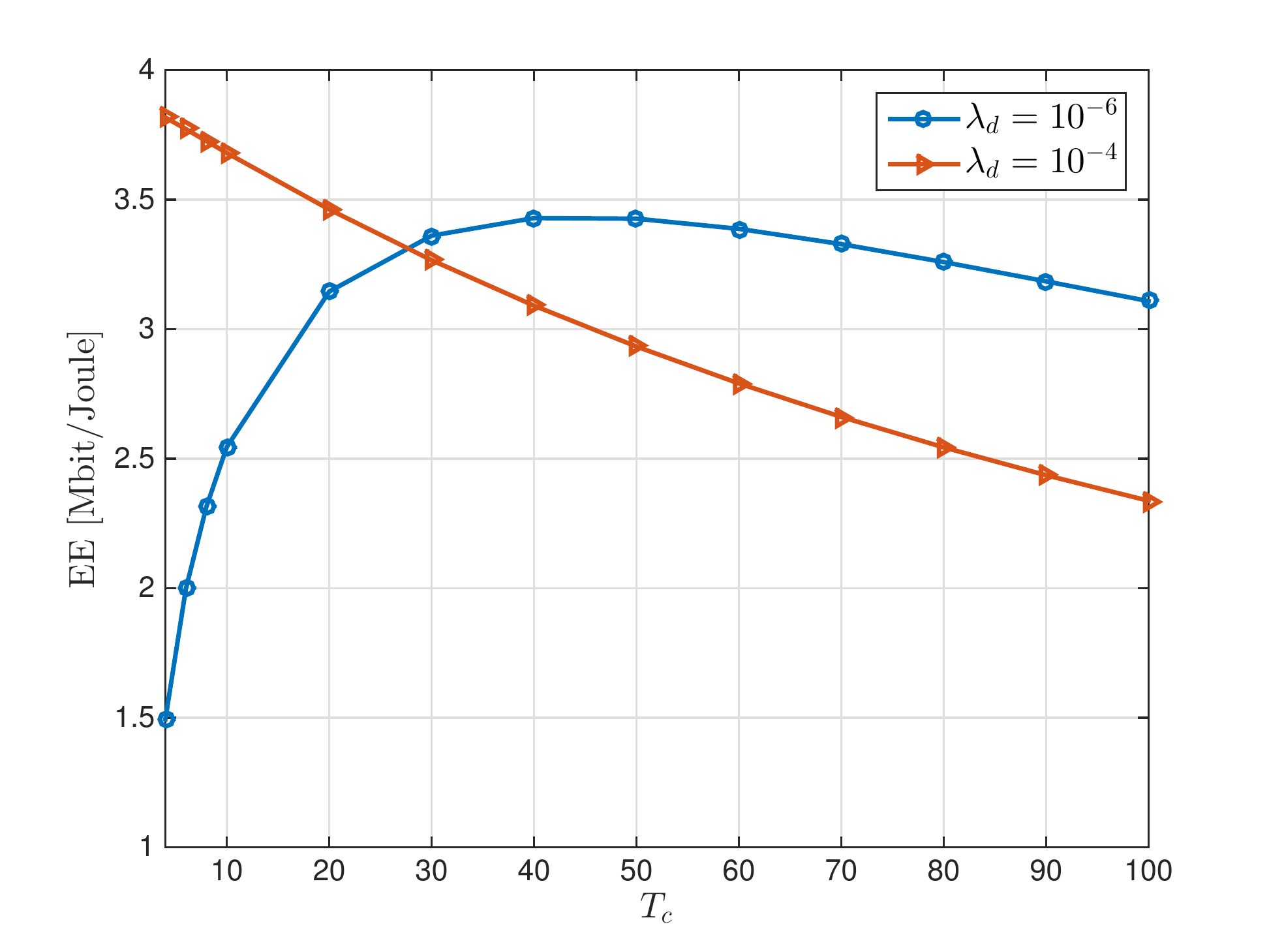}
\vspace*{-3mm}
\caption{EE $\mathrm{[Mbit/Joule]}$ as a function of BS antennas $T_c \in \{4,\dots,100\}$ for $\lambda_d=\{10^{-6},10^{-4}\}$.}
\label{fig:EE_ld-4-6_tc}
\vspace*{-6mm}
\end{figure}

\figref{fig:EE_ld_tc_uc4} shows the network performance in terms of EE versus the parameters $\lambda_d$ and $T_c$. To study this results further, similar to the ASR, we first plot the EE versus $\lambda_d$ for $T_c = \{4,70\}$ in \figref{fig:EE_ld_tc4_tc70}. We can see that the pattern for both lower and higher number of antenna is similar to \figref{fig:ASR_ld_tc4_tc70}. However, if we plot the EE versus $T_c$, we see a different behavior for low- and high-density D2D scenario. From \figref{fig:EE_ld-4-6_tc}, it is observed that indeed the low-density scenario benefits in terms of the number of antennas until the sum of the circuit powers for all antennas dominates the performance and leads to a gradual decrease in EE. As the figure implies, there exists an optimal number of antennas ($T_c$) for maximal EE in the low-density scenario. However, in high-density D2D scenario, which is an interference-limited scenario, the EE decreases almost linearly with $T_c$, as depicted in \figref{fig:ASR_ld_tc4_tc70}. Increasing the number of antennas in this region cannot improve the ASR much as shown in \figref{fig:ASR_ld-4-6_tc}. Then, at the same time the circuit power resulting from higher number of antennas increases which in turn leads to poor network EE.

\section{Conclusions}
\label{sec:conclusion}
We studied the coexistence of two key 5G concepts: device-to-device (D2D) communication and massive MIMO. We considered two performance metrics, namely, average sum rate (ASR) in $\mathrm{bit/s}$ and energy efficiency (EE) in $\mathrm{bit/Joule}$. We assumed that there is a fixed number of randomly distributed  cellular user equipments (CUEs) in the network, while the number of D2D transmitters are distributed according to a Poisson point process (PPP). We derived tractable expressions for ASR and EE, and studied the tradeoff between the number of base station antennas and the density of D2D users with a fixed number of CUE for a given coverage area in the downlink. Our results showed that both ASR and EE have different behaviors in scenarios with low and high density of D2D users. By increasing the number of antennas in the low D2D density regime, the ASR improves, however, EE increases until the circuit power from many antennas becomes dominant. In the high D2D density regime, there is small gain in terms of the ASR from adding many antennas to the detriment of EE which decreases tremendously.

\vspace*{-3mm}

\bibliographystyle{IEEEtran}
\bibliography{IEEEabrv,refs,serveh}

\end{document}